# Smooth Manifolds of Kinematic Type.

*V. R. Krym*

**Operations Theory, Mathematics Department of the St.-Petersburg State University, Bibliotechnaya pl. 2, Peterhof, St.-Petersburg 198904 Russia.**

E-mail: *vkrym@pdmi.ras.ru*

We give a short historical review of early Kaluza-Klein theories. We study various causal structures on manifolds, especially those which cannot be described by a metric tensor with signature (+---). The smooth structure (atlas) on a manifold is found to be related with its causal structure.

## §1. Introduction.

**1.1.** This paper is devoted to relations expressing «causality» on smooth manifolds. Interest to this direction was provoked mainly by the special and general relativity, yet it was not all of the causal structures considered up to the present time. In this paper we review different types of causality on a smooth manifold. We prove also that there is a clear relation between the causal structure and the smooth structure on a manifold. In the next paper we shall use some special smooth structure on a 5-manifold for unification of the electromagnetic and gravitational interactions.

**1.2.** It is well known that all early attempts to unify the fundamental interactions (including gravity) faced serious problems. The paper by Kaluza [1], historically first on this subject, was approximate in origin. In 1926 V.A.Fock [2] proved that the trajectory of motion of a charged particle can be strictly described as a geodesic line of zero length on a 5-manifold. Independently O.Klein [3] used the same metric tensor on a 5-manifold to deduce the 4-dimensional Einstein's and Maxwell's equations. Klein did not study the geodesic equations, but noted only that the equations of motion follow as usual from the Einstein's equations. The authors of these papers implied that the metric tensor had the signature (+----), i.e. the fifth coordinate is spacelike. It was also assumed that the metric tensor did not depend on the fifth coordinate and $g_{55}=1$. Both the conditions were not invariant.

In 1956 Yu.B.Rumer [4] gave up the condition $g_{55}=1$. The fifth coordinate was also considered to be spacelike. Rumer deduced the classical equations of motion of a charged particle, but he (as well as Fock) also used the assumption that on a 5-manifold they are equations of geodesics of zero length, i.e. lightlike. This feature cannot be completely understood. Rumer also wrote Einstein's equations for a 5-manifold with his own metric tensor. Unfortunately this system of differential equations was not equivalent to the classical Maxwell's and Einstein's equations in the presence of electromagnetic field.

Einstein and Bergman in 1938 [5] supposed that space-time is cyclically closed in the direction of the fifth coordinate. Their purpose was to explain why all observable physical quantities do not depend on the fifth coordinate. According to Klein and Fock, the period in this direction can be assumed to be $\hbar/mc$. Rumer also considered space-time to be closed as a cylinder in this direction, but with the period $\hbar$. Yet the fifth coordinate according to Rumer was the action (with the same dimension), and to claim the action to be cyclic is incorrect because in that case the principle of the minimal action loses any sense. Also, action is not an independent variable. Five dimensional space-time was considered by Rumer as a configuration space for a given particle, and his metric tensor depended on the ratio of its charge to its mass. Yet if every particle had being moved in its own configuration space, and common for all particles space-time had had no physical meaning, then how particles could exchange interactions with each other?

In 1960s Busemann [6] and Pimenov [7,8] concluded that the unified theory of gravitational and electromagnetic interactions could not be constructed without previous study of the causality on manifolds. Busemann and Pimenov independently introduced similar sets of axioms and studied the ordering on manifolds. Pimenov wrote that the additional coordinate could not be either timelike or spacelike. "The usual Riemannian geometry is well suited to simulate the values of two types only: real and imaginary. In physics it corresponds to simulation of only two physical dimensions. Yet the theory of electromagnetism, besides dimension «time» and dimension «space», deals with at least one more dimension «electricity». Therefore the theory of electromagnetism and non-degenerated Riemannian geometry are not homologic in general sense of word" [7, p. 424].

Using these principles Pimenov created a unified theory of gravitational and electromagnetic interactions. The fifth coordinate was independent, as it should be in geometry. Yet Pimenov could not deduce the classical equations of motion of a charged particle. The principle of the minimal action was not employed in his paper. In the Einstein-like equations Pimenov introduced a very formal object which was not the Ricci tensor in any of the two connections considered in his work. The system of Einstein's and Maxwell's equations could be deduced so, yet such a deduction violated traditions of mathematical physics.

In this paper we prove that Pimenov's kinematic types naturally lead to specific smooth structures on a manifold, which have not been considered earlier. These structures are extremely useful in unifications of fundamental interactions. It will be the subject of the next paper.

## §2. Smooth manifolds with causal relations.

Let us start with a linear space with some topology $\Delta$. In a linear space the (partial) ordering $<$ should be chosen to be compatible with the linear structure and the topological structure on this space [9,10]. Such an ordering can be fully described by an appropriate cone. In the linear space $L$ (which can be infinite-dimensional) the set of all points which follow the given point $a \in L$ is a convex cone $Q_a^+$ with vertex $a$, $a \notin Q_a^+$. The set $Q_0^+$ of points which follow the null $0 \in L$ is called the <u>future cone of the kinematics</u> $L$. For any point $a \in L$ $Q_a^+ = a + Q_0^+$. The set of all points which precede the given point $a \in L$ is $Q_a^-$, $Q_0^- = -Q_0^+$ and $\forall a \in L$ $Q_a^- = a + Q_0^-$.

The topology $\Delta$ of a linear space $L$ is chosen so that the future cone $Q_0^+$ be opened. The space $L$ is not «linear topological» in the usual sense because we do not require the multiplication mapping $L \times \mathrm{R} \to L$ be continuous. Instead of it we require that the mapping of multiplication on every fixed scalar be continuous [9,10]. If the space $L$ is finite dimensional, the topology in $L$ can be explicitly related with the type of the future cone. If the future cone contains lines, then the weakest topology in $L$ is indiscrete on every line which is contained in the future cone. If the affine dimension of the future cone is lower than the dimension of $L$, then the topology in $L$ contains «discrete layers». If the future cone is sharp (i.e. it does not contain lines), has the same dimension as $L$, and its boundary does not contain rays, then the weakest topology in $L$ is Euclidean.

Transfer the relation of ordering from the finite-dimensional linear kinematics ($L$, $<$, $\Delta$) to the smooth manifold $M$. The mapping $\varphi:L\to L$ <u>preserves ordering</u> in $L$ iff for any $x,y\in L$  $x<y \Leftrightarrow \varphi(x)<\varphi(y)$.

**2.1. Definition.** *Let $M$ be a smooth manifold, $L$ be a linear kinematics of the same dimension. Let $P$ be a family of continuous mappings $(P(U))_{U\in I}$, $P(U):TU\to L$ $\forall U\in I$, where each element $U\in I$ is a connected open subset (domain) of $M$ such that*

*1. $\bigcup\limits_{U\in I} U = M$ (i.e. $I$ is a covering of $M$);*

*2. $\forall U\in I$ $\forall x\in U$ the restriction $P(U)_x:T_xU\to L$ is a non-degenerate linear mapping;*

*3. $\forall U,V\in I$ $\forall x\in U\cap V$ the composition mapping $P(U)_x \circ P(V)_x^{-1}:L\to L$ preserves ordering in $L$.*

*Then $M$ with the indicated family $P$ is called the <u>manifold of the kinematic type</u> $L$. The family[1] $P$ is called the <u>kinematic atlas</u> on $M$.*

In any tangent space $T_aM$ of the manifold of the kinematic type $L$ the cone $P(U)_a^{-1}(Q_0^+)$ is correctly defined. It is the pre-image of the future cone of the null point of the kinematics $L$ with the mapping $P(U)_a$. Due to (3) this cone does not depend on the domain $U$ $(a\in U)$, therefore we drop the notation $U$. Define $Q_a^+ := P_a^{-1}(Q_0^+)$, $Q_a^- := P_a^{-1}(Q_0^-)$. The family of the cones $Q_a^+$ (and $Q_a^-$) continuously depends on the point $a\in M$ in the sense that the mapping $P(U)$ is continuous.

**Example 1.** Each linear kinematics $L$ is a manifold of the kinematic type $L$. Its kinematic atlas consists of the single mapping $P_x:=id_L-x$ $\forall x\in L$. The ordering (i.e. the future cone $Q_0^+$) in $L$ should fulfill several natural axioms [9,10]. The Newton's ordering is defined by the condition $t>0$, i.e. the future cone is the half-space. The Einstein's ordering is defined by an elliptic cone. The weakest topologies are different for these two orderings: for the Newton's causality the natural topology is $R^1\times R_a^{n-1}$, where $R_a$ is the real line with the indiscrete topology, $n:=dim\ L$, and for the Einstein's causality the weakest topology is Euclidean. There are also other possible types of causality.

---

[1] These mappings are not differentials of some charts on $M$, in general. If all these mappings are differentials of some charts, then for elliptic cones due to (3) there is a linear structure on the manifold $M$, and that is (in a sense) strange.

**Example 2.** Let $M$ be a parallelizable manifold of dimension $n$, $X_1,...,X_n$ be continuous vector fields on $M$ linearly independent in any point. Let $L$ be a linear kinematics of the same dimension. Define the mapping $P:TM \to L$, $P_a(Y):=(Y^1,...,Y^n)$, where $Y=:\sum_{k=1}^{n} Y^k X_k(a)$. The couple $(M, P)$ is a manifold of the kinematic type $L$.

**2.2. The orientation of time.**

**Definition.** *Let $(M, P)$ be a manifold of the kinematic type $L$. The <u>orientation of time</u> of the manifold $M$ is a continuous vector field $E$ on $M$ so that $\forall a \in M\ \ E(a) \in Q_a^+$.* In particular the field $E$ never turns to be zero.

**Theorem 1.** *On any manifold $(M, P)$ of the kinematic type $L$ there is the orientation of time.*

**Proof.** Let us choose a vector $v \in Q_0^+ \subset L$. For any point $a \in M$ there is a neighbourhood $U$ where the mapping $P(U)$ of the family $P$ is defined. Therefore there is a continuous vector field $Y_U$, $\forall x \in U\ \ Y_U(x):=P(U)_x^{-1}(v)$. Let $(\chi_U)_{U \in I}$ be the partition of unity associated with the covering $I$. Since the manifold $M$ is locally compact, the covering $I$ can be taken locally finite. Since the cone $Q_a^+$ is convex, the vector field $E:=\sum_{U \in I} \chi_U Y_U$ is the required vector field (at each point this sum contains only finite number of non-zero terms).

**2.3.** Let us discuss typical examples of composition mappings $P(U)_x \circ P(V)_x^{-1}$ for various linear kinematics $L$.

**2.3.1.** Let the future cone $Q_0^+ \subset L$ be elliptic. Then due to well known theorems each linear mapping which preserves the cone $Q_0^+$ is the Lorenz transformation with some arbitrary positive factor. In other words, linear mappings preserving $Q_0^+$ are hyperbolic rotations in the pseudometric of signature (+-...-), which can be uniquely constructed based on the cone $Q_0^+$ (disregarding scaling factor).

**2.3.2.** Let the future cone $Q_0^+$ contain al least one line. Then [9, 10] the weakest topology of the kinematics $L$ contains non-trivial linear indiscrete subspace. Consider the maximal linear subspace $l_0$ containing in the closure of the future cone $Q_0^+$ of the kinematics $L$. Let $n:=dim\ L,\ k:=dim\ l_0$. Let $e_{n-k+1},\ ...,\ e_n$ be a basis of the subspace $l_0$. The subspace $l_0$ is transferred into itself by all linear mappings which preserves ordering. Therefore in the coordinate system where $x^j$-coordinate is defined as the coefficient at $e_j$ ($j=n-k+1, ..., n$) the matrix of the composition mapping $P(U)_x \circ P(V)_x^{-1}$ has the block structure
$$\begin{pmatrix} & * & & \begin{matrix} 0 \\ \vdots \\ 0 \end{matrix} \\ \hline * & \cdots & * & * \end{pmatrix}$$

The left upper block (square of dimension $n-k$) is a non-degenerate linear mapping. The right upper block is zero. The right down block (square of dimension $k$) is a non-degenerate linear mapping also. The set of these matrices is a group.

**2.3.3.** Let the dimension of the future cone $Q_0^+$ of the kinematics $L$ be lower than dimension $L$. Then [9, 10] the weakest topology of the kinematics $L$ contains a non-trivial linear discrete subspace. The minimal linear subspace $l_0$ that contains the future cone $Q_0^+$ is transferred into itself by all linear mappings which preserves ordering. Let $n:=dim\ L$, $k:=dim\ l_0$, and let $e_1, ..., e_k$ be the basis of the subspace $l_0$. Then the matrix of the composition mapping $P(U)_x \circ P(V)_x^{-1}$ has the block structure
$$\begin{pmatrix} & * & & \begin{matrix} * \\ \vdots \\ * \end{matrix} \\ \hline 0 & \cdots & 0 & * \end{pmatrix}$$

The left upper block (square of dimension $k$) is a non-degenerate linear mapping. The left down block is zero. The right down block (square of dimension $n-k$) is a non-degenerate linear mapping also. The set of these matrices is a group.

**2.3.4.** The set of linear isomorphisms $L \to L$ which preserves ordering in $L$ is a group. We designate this group as $GL(L, <)$. There can be such kinematic atlases $(P(U))_{U \in I}$ on the manifold $M$ that the mentioned composition mappings belong to some subgroup $H$ of this group. For example the parallelizable manifold (Example 2) has a kinematic atlas with the single mapping, the subgroup $H$ hence being unit. The inverse statement is also true: if a manifold has a kinematic atlas such that all its composition mappings are identities, then this manifold is parallelizable.

The set of linear isomorphisms $L \to L$ which preserves linear subspaces $l_0$ defined in 2.3.2 and 2.3.3 is designated as $GL(L, Lin(<))$. It is a subgroup of $GL(L, <)$.

### 2.4. Local ordering.

**2.4.1. Definition.** *Let (M, P) be a manifold of the kinematic type L, $\gamma:[a,b] \to M$ be a piecewice smooth path. The path $\gamma$ is called <u>directed in future</u> if $\forall t \in [a,b]$ (where derivative is defined) $\gamma'(t) \in Q^+_{\gamma(t)}$. If the path $\gamma$ is not differentiable at a point t, then both the derivatives on the left and on the right should satisfy this condition at t. The path <u>directed in past</u> is defined similarly. The path $\gamma$ is called <u>timelike</u> if it is directed in future or in past.*

**Definition.** *Let (M, P) be a manifold of the kinematic type L, $U \subset M$, points $x,y \in U$, and there is a path $\gamma:[a,b] \to U$ directed in future with the beginning point x and ending point y (i.e.. $\gamma(a)=x$, $\gamma(b)=y$). Then we designate <u>x<y (mod $\gamma$)</u>. Similarly if the path $\gamma$ is directed in past, then designate <u>x>y (mod $\gamma$)</u>.*

Obvious that $x<y$ (mod $\gamma$) $\Leftrightarrow$ $y>x$ (mod $\gamma^{-1}$).

One would like to interpret < as a local ordering in *M*, yet it is not always possible.

**2.4.2. Example 3.** Let $U \subset M$ be an open connected set and *(M, P)* be a manifold of the kinematic type *L*. Let also the closure of the future cone of *L* contain a non-trivial linear subspace $l_0$. The future cone of *L* is supposed to be «wing-type», that is dim aff $(L\setminus Cl(Q^+_0 \cup Q^-_0))$=dim *L*. Then there is a distribution (a field of linear subspaces) $l_x := P_x^{-1}(l_0)$ $\forall x \in M$. Assume that the distribution $(l_x)_{x \in U}$ is smooth, completely non-holonomic and strongly generating. Then there are smooth vector fields *X, Y* on *U* such that $X(a), Y(a) \in l_a$ $\forall a \in U$, but $[X,Y]_a \notin Cl Q^+_a$ $\forall a \in U$ (the domain *U* may be made smaller with the same notation). It is well known that motion along the integral curve of the vector field *[X, Y]* can be realized with subsequential shifts along the integral curves of the vector fields *X, Y*. Since $X(a), Y(a) \in Cl Q^+_a$ $\forall a \in U$, then motion along these integral curves can be approximated as exactly as needed by motion along timelike paths in *U* directed in future. Then for any points $a,b \in U$ and for any neighbourhood *V* of the point *b* there is a point $c \in V$ such that $\exists \gamma$ *a<c (mod $\gamma$)*, no matter how tiny is the set *U*.

**Theorem 2.** *Let (M, P) be a manifold of the kinematic type L. Let the future cone of L be such that* dim aff $(L\setminus Cl(Q^+_0 \cup Q^-_0))$=dim *L. Let $l_0$ be the maximal linear subspace containing in the closure of the future cone of L. Let the distribution* $l_x := P_x^{-1}(l_0)$, $x \in M$, *be holonomic. Then for each point $x \in M$ there is a neighbourhood U such that any timelike path in U has no self-intersection points.*

If the conditions of Theorem 2 are fulfilled and domain $U$ is as indicated, then we use notation <u>*a<b (mod U)*</u>. The relation $<$ in these conditions does not depend on the path $\gamma$, depending only on the domain $U$.

**Proof.** For sufficiently small domain $U$ the holonomic distribution $l$ can be expanded up to the distribution $s$ so that

1). $\forall x \in U$  $l_x \subset s_x$

2). Distribution $s$ has codimension 1 and holonomic.

3). $\forall x \in U$  $s_x$ separates $Q_x^+$ from $Q_x^-$.

Let us first fix a point $x \in U$. The cones $Q_x^+$ and $Q_x^-$ are convex, they do not intersect and therefore they can be separated by a hyperplane $s_x$. Since dim aff $(L \backslash \text{Cl}(Q_0^+ \cup Q_0^-)) = \dim L$, this hyperplane can be chosen so that $s_x \cap \text{Cl}(Q_x^+ \cup Q_x^-) = l_x$ (obvious that $l_x \subset s_x$). The hyperplane $s_x$ is the tangent space to a smooth submanifold of codimension 1 at the point $x$. This submanifold can be chosen so that it contains the integral submanifold of distribution $l$ which passes through the point $x$. Consider the distribution $s$ of the tangent spaces to the above submanifold of codimension 1. Since the families of cones $\left(Q_x^+\right)_{x \in U}$ and $\left(Q_x^-\right)_{x \in U}$ continuously depend on the point $x$, therefore for sufficiently small domain $U$ the hyperplanes $s_x$ separates the cones $Q_x^+$ and $Q_x^-$ for any $x \in U$.

Let $\gamma$ be a path in $U$ directed in future. Consider the maximal on $U$ integral submanifold of the distribution $s$, which passes through the point $x \in \text{im } \gamma$. If the domain $U$ is sufficiently small then this submanifold separates $U$ in two connected components. The path $\gamma$ starts up in one of these components and ends up in the other. Therefore there are no closed timelike paths in $U$.

In the following we consider only such manifolds of the kinematic type $(M,P,L)$ that satisfy the conditions of Theorem 2.

**2.4.3. Properties** of the relation $<$.

1). *The relation $<$ is topologically non-empty, i.e. in any neighbourhood U of any point $a \in M$ there are points $x, y \in U$ such that $x < y$ (mod U).*

2). *The relation $<$ is antireflexive, i.e. for any point $x \in M$ there is a neighbourhood $U \ni x$ so that $\neg(x < x$ (mod U)).*

3). *The relation < is locally transitive, i.e. for any point a∈M there is a neighbourhood U such that $\forall x,y,z \in U$ x<y (mod U) & y<z (mod U) $\Rightarrow$ x<z (mod U).*

**Definition.** *Let M be a manifold, < be a relation on M with properties (1-3). Then < is called the <u>local ordering</u> on M.*

Then, for each manifold *(M, P)* of the kinematic type *L* a local ordering is defined. Local ordering depends on the ordering in *L* and the kinematic atlas *P*. We shall not consider the global transitive closure [11] of the relation <, since this closure may not be an ordering (even locally). For example, local ordering can be defined on a circle, while global ordering cannot.

**2.4.4.** Local ordering < defines a local equivalence on the manifold.

**Definition.** *Let $x,y \in M$, $U \ni x$ be a domain on the manifold M which satisfies the conditions of the theorem 2. Designate $K_x^+ := \{a \in U \mid a > x \ (mod \ U)\}$, $K_x^- := \{a \in U \mid a < x \ (mod \ U)\}$. Introduce the relations*

$$\underline{x \leq y \ (mod \ U)} \ \Leftrightarrow \ x \in Cl \ K_y^-$$

$$\underline{x \approx y \ (mod \ U)} \ \Leftrightarrow \ x \leq y \ (mod \ U) \ \& \ y \leq x \ (mod \ U)$$

**Properties** of the relation ≈.

1. *Reflexivity* follows from the fact that in any neighbourhood of the given point there are points of the future and of the past in the sense of local ordering <. The relation ≤ is also reflexive.

2. ≈ is *symmetric* by definition.

3. *Transitivity* follows from the transitivity of ≤.

In the classical special relativity the future cone is elliptic and the relation ≈ is trivial: the class *x/≈* of equivalence of any point *x* consists of this point only. In the Newton's kinematics the future cone is the half-space and the relation ≈ is understood as a simultaneity. The class *x/≈* for any point *x* is a hyperplane.

**2.4.5.** It is well known that each cone that contains lines is a Cartesian product of a sharp cone and an affine subspace. Then if the future cone of the kinematics *L* contains a line, then there is a distribution of dimension *m* on the manifold *M* of the kinematic type *L,* where *m* is the dimension of the maximal linear subspace contained in the closure of the future cone. Let the domain *U* satisfy the conditions of the theorem 2. Then for any point $x \in U$ the class *x/≈* of equivalence ≈ is a submanifold closed in *U*. It is exactly the maximal (in $U$) integral manifold of the mentioned distribution that passes through the point $x$. Topological space $U/\approx$ is a manifold.

Designate $\sim$ the transitive closure of the relation $\approx$ on the whole $M$. The class $x/\sim$ may not be closed in $M$, and factorspace $M/\sim$ may not be a manifold even for an one-dimensional distribution.

**Example 4.** Let $M$ be a cylinder $S^1 \times R$. Define the mapping $P$ on $M$:

$$(u,v) \in T_{(\alpha,x)}M \quad \mapsto \quad P_{(\alpha,x)}(u,v) := \begin{cases} \begin{pmatrix} u \\ v \end{pmatrix}, & (x \geq x_0) \\ \begin{pmatrix} 1 & x-x_0 \\ x_0-x & 1 \end{pmatrix}^{-1} \begin{pmatrix} u \\ v \end{pmatrix}, & (x \leq x_0) \end{cases}$$

($x$ is the height along the generatrix of the cylinder, $\alpha$ is the coordinate on the circuit).

Let us choose two-dimensional linear kinematics $L$ with the cone $Q_0^+$ as the open half-plane. The couple $(M, P)$ is a manifold of the kinematic type $L$. To find classes $x/\approx$ of equivalence $\approx$, let us find preimage of the indiscrete fiber of the kinematics $L$ at the mapping $P$:
$$\begin{pmatrix} 1 & x-x_0 \\ x_0-x & 1 \end{pmatrix}\begin{pmatrix} 1 \\ 0 \end{pmatrix} = \begin{pmatrix} 1 \\ x_0-x \end{pmatrix}$$

This is a tangent vector field on $M$. Let us integrate it. The system of differential equations $\dot{\alpha}=1$, $\dot{x}=x_0-x$, has the general solution $\alpha=t+c_1$, $x=x_0-e^{-t+c}$. Therefore the classes of relation $\sim$ before the height $x_0$ are spirals with infinite number of rotations around the cylinder, and after the height $x_0$ they are circuits. The factorspace $M/\sim$ consists of a circuit and a ray containing its vertex. Any neighbourhood of the ray's vertex contains the whole circuit. This topological space is not a manifold and it cannot be imbedded into $M$ (Fig. 1 left).

**Example 5.** Cylinder with a cut-off can be also turned into a manifold of kinematic type $R \times R_a$ (Fig. 1 right). Let the mapping $P$ be the same as in Example 4 at $x \geq x_0$. Then factorspace is just $R$ with the usual ordering, but in general the classes of equivalence $\sim$ of different points are not homeomorphic.

**2.5. Smooth mappings of manifolds of the kinematic type.** Let $N$ be a smooth manifold, $(M, P)$ be a manifold of the kinematic type $L$, $\varphi:N \to M$ be a smooth mapping, dim $M$=dim $N$. Then there is a family $(P(U) \circ d\varphi)_{U \in I}$ on $N$. It satisfies the conditions of the Definition 2.1, iff at each point $x \in N$ the differential $d_x\varphi$ is non-degenerated and $P(U) \circ d\varphi \circ (P(V) \circ d\varphi)^{-1}$ preserves ordering in $L$ for any (non-empty) $U \cap V$. Then $(im\ \varphi, P \circ d\varphi)$ is also a manifold of the kinematic type $L$. For example, the real line R with the usual ordering can be projected onto $S^1$ defining kinematic structure on it.

**Definition.** *Let $(M, P)$ and $(N, T)$ be smooth manifolds of the kinematic type L, $P=(P(U))_{U \in I}$, $T=(T(V))_{V \in I'}$. Smooth mapping $\varphi: N \to M$ is called a <u>diffeomorphism of manifolds of the kinematic type</u> if $\varphi$ is a diffeomorphisms of N and M as smooth manifolds and*

$$\forall U \in I \ \forall V \in I' \ \forall x \in V \ \varphi(x) \in U \Rightarrow (P(U)_{\varphi(x)} \circ d_x\varphi)^{-1}(Q_0^+) = (T(V)_x)^{-1}(Q_0^+)$$

Hence the mappings $P(U)_{\varphi(x)} \circ d_x\varphi \circ (T(V)_x)^{-1}$ are linear automorphisms of the kinematics $L$ onto itself that preserve ordering in $L$. The relations of local ordering on the diffeomorphic manifolds of the kinematic types are obviously the same (if these manifolds are identified with the diffeomorphism $\varphi$). Manifolds $N$ and $M$ will be called <u>diffeomorphic with respect to ordering</u>.

**2.6. Ordering of the cotangent space.** We have the relation of local ordering on the manifold $M$ and ordering on any tangent space $T_xM$. We shall need also ordering of the cotangent space. Define $\forall \omega \in T_x^*M \ \ \omega > 0 \Leftrightarrow \forall u \in Q_x^+ \ \ \omega \cdot u > 0$ (action of a covector on a vector is designated by a dot). In other words, ordering in the cotangent space is defined by a cone which is dual to the future cone of the tangent space. There is a topology on $T_x^*M$ for which this space is a linear kinematics. In general the kinematic type changes during transition from the tangent space to its dual: indiscrete fiber forces discrete fiberization and vice versa. If the weakest topology in the kinematics $L$ is Euclidean, the weakest topology in $L^*$ is also Euclidean.

## §3. Pseudo-Riemannian manifolds.

**3.1.** Each pseudo-Riemannian manifold $M$ with an inner product $< , >$ of signature **(+-...-)** and orientation of time $E$ (a continuous timelike vector field which is never zero) has some canonical causality. It is given by the kinematic atlas defined as followed.

For each point $a \in M$ there is a neighbourhood $U$ where geodesics[1] with the starting point $a$ do not intersect. Let $P_x(v)$ be a vector obtained by the parallel transfer of the vector $v \in T_xM$ along the geodesic path which connects points $x$ and $a$ in $U$ to the point $a$. The mapping $P$ is continuous and non-degenerate.

---

[1] Defined in some Riemannian (signature (+...+)) metrics which exists on $U$.

Define $Q_x^+ := \{v \in T_x M \mid <v, v> > 0 \ \& \ <E(x), v> > 0\}$.

**Lemma.** *Let connection $\nabla$ on $M$ be (pseudo)Riemannian (i.e. $X<Y,Z> = <\nabla_X Y, Z> + <Y, \nabla_X Z>$ for any vector fields X,Y,Z). Then the parallel transfer P along each path $\gamma$ preserves ordering: $P_x(Q_x^+) = Q_a^+$* [14].

The tangent space $T_a M$ in any point $a \in M$ can be chosen as a linear kinematics *L*, since all of them have the same ordering. Obvious that the cone $Q_a^+ \subset L$ is elliptic. Therefore for each point $a \in M$ there is a neighbourhood *U* where there is the mapping *P* (defined above). The couple (*U, P*) is a manifold of the kinematic type *L*. If two such domains intersect, then the appropriate mappings are compatible with each other: $\forall x \in U \cap V \ \ P(U)_x^{-1}(Q_0^+) = P(V)_x^{-1}(Q_0^+)$. We proved

**Theorem 3.** *Let M be a pseudo-Riemannian manifold with an inner product < , > of signature ( +-...- ) and let E be the orientation of time on M. Let P be a family of all mappings constructed above. Then (M, P) is a manifold of the kinematic type L, where L is any of the tangent spaces $T_a M$ with the restriction of pseudo-Riemannian metric on it, the future cone of this kinematics being chosen so that it contain vector E(a).*

Riemannian connection mentioned in the Lemma above is not unique. One can also change the family of curves which are used for parallel transfer. Different kinematic atlases *P* can be constructed in this way. Yet all manifolds *(M, P)* will be diffeomorphic as manifolds of the kinematic type *L*.

**3.2. Problem.** *When based on the given manifold (M, P) of the kinematic type L one can construct a kinematic atlas Z such that all mappings $Z(U): TU \to L$ are differentials of some smooth charts $h: U \to L$, and (M, P) is diffeomorphic (M, Z) with respect to ordering?* (See also [7, p. 351, 416]).

This problem probably cannot be solved positively on every manifold. Yet the existence of the kinematic atlas does not restrict manifold as a whole too strongly.

**Example 6.** Kinematic ordering can be introduced on the non-orientable manifold, e.g., on the Möbius sheet. Map the future cone of a two-dimensional linear kinematics *L* into *TM* so that the symmetry axis of the cone was parallel to the boundary of the Möbius sheet. The weakest topology in *L* can be both $R \times R_a$ and $R^2$.

## §4. Pseudometric smooth manifolds.

**4.1.** In this paragraph we prove that the requirements to the pseudo-Riemannian metrics can be weakened.

**Definition.** *Let M be a smooth manifold. M is called a <u>pseudometric manifold</u> if there is a symmetrical bilinear form $< , >_x$: $T_xM \times T_xM \to$ R of signature (+-...-0...0) defined on TM that smoothly depends on the point $x \in M$. The form $< , >$ is called an inner product.*

We consider only such pseudometric manifolds that have the orientation of time, i.e. the continuous vector filed $E$ so that $<E, E> > 0$ on the whole manifold. Define $F_x := \{u \in T_xM |\ <u, u> > 0\} \cup \{0\}$ and a function on it: $f_x : F_x \to$ R: $f_x(u) := sgn(<E(x), u>) \cdot <u, u>^{1/2}$ ($sgn(0) := 0$), $x \in M$. The future cone of the appropriate linear kinematics is defined naturally as $\forall u \in T_xM\ \ u \in Q_x^+ \Leftrightarrow u \in F_x$ & $f_x(u) > 0$. The function $f_x$ has the following properties:

1). $\forall u, v \in Q_x^+\ \ <u,v> \geq f_x(u) f_x(v)$

2). $\forall u, v \in Q_x^+\ \ f_x(u+v) \geq f_x(u) + f_x(v)$

3). $\forall u \in F_x\ \ \forall \alpha \in$ R $\ \ f_x(\alpha u) = \alpha f_x(u)$

In other words the couple $(T_xM, f_x)$ is a pseudometric linear kinematics according to the Definition 5.1 from [9, 10].

Since the signature of the inner product does not depend on the point on $M$, all these kinematics $(T_xM, f_x)$ are linearly isomorphic with respect to ordering. Designate any of these kinematics as $L$.

We constructed a continuous family $(Q_x^+)_{x \in M}$ of identical cones on $M$. Prove that this family can be defined by some kinematic atlas.

Let us choose $n$ linearly independent continuous vector fields $(X_k)_{k=1,...,n}$ in some domain $U$, $n := dim\ M$. Transfer the matrix of the metric tensor in each tangent space $T_xM$, $x \in U$, to the canonical form $Diag(1, -1, ..., -1, 0, ..., 0) = V_x^T g_x V_x$. The linear operator $V_x$ is not unique. In the maximal subspace where the metric tensor has the signature (-...-) the mapping $V_x$ is defined with an arbitrary orthogonal linear mapping. In the maximal subspace where the metric tensor is degenerated the mapping $V_x$ is defined with an arbitrary non-degenerate linear mapping. Therefore the basises in these subspaces should be chosen in a more definite way.

All vectors in each tangent space $T_xM$ we write as linear combinations of the previously chosen $(X_k(x))_{k=1,...,n}$. Let $(X_k(x))^T_{k=1,...,n} V_x$ be the basis of $T_xM$ (in this basis the matrix of the metric tensor is already diagonal). Consider the maximal subspace where the metric tensor has the signature (-...-). Let the dimension of this subspace be *m*. Basis vectors of this subspace are written as columns of coordinates. The rank of this rectangular matrix is *m*. Choose a non-degenerate square matrix of order *m* from it. The domain *U* can be taken so small that this submatrix have the maximal rank on the whole *U*. Let us choose the basis of the subspace in question so that the upper right part of this submatrix (above its diagonal) be zero. The dimension of the group of orthogonal transformations *O(m)* as a smooth manifold is *m(m-1)/2,* and we impose the same number of independent conditions on the basis vectors. To define the linear mapping $V_x$ uniquely, choose the phase factors ±1 of the basis vectors. They can be defined by the condition <*E(x), u*> >0 (if some of the basis vectors prove to be orthogonal to *E(x)*, then the orientation of time *E* should be changed unessential so that the field *E* and this set of the basis vectors were not orthogonal on *U*. Change of the orientation of time is called unessential if new orientation of time lies in the same connected component of the cone $F_x$|*{0}* for all *x*∈*M*).

In the maximal subspace where the metric tensor is degenerated we impose $m^2$ conditions on the basis vectors since the dimension of the group *GL(m)* as a smooth manifold is $m^2$. We require that the coordinates of the basis vectors form a submatrix (chosen as above) equal to the unit diagonal matrix. Phase factors are defined by this requirement too.

Tangent spaces in these new sets of the basis vectors can be identified with each other. The mapping *V* defined in the each point *x*∈*U* by the matrix $V_x$ is the required. In general this vector-type 1-form *V* is not closed. Since the domain *U* is arbitrary (the whole *M* can be covered by such domains), the kinematic atlas *P={V(U)|U*∈*I}* is constructed. The composition mappings satisfy the condition (3) of the Definition of the manifold of kinematic type *L* (in particular, the condition 2.3.2). Thus we proved

**Theorem 4.** *Let M be a pseudometric manifold with an inner product < , > of signature (+-...-0...0) and an orientation of time E. Let P be the kinematic atlas constructed above. Then (M, P) is a manifold of the kinematic type L, where any tangent space $T_aM$ with the restriction of pseudometrics can be taken as L. The future cone of this kinematics is chosen so that it contain vector E(a).*

**4.2. Coconnection on a pseudometric manifold.**

**Definition.** *Let M be a smooth manifold with the inner product $< , >$, $a \in M$. A rule $\overset{*}{\nabla}$ which maps each vector $u \in T_a M$ and each smooth vector field $X$ to a covector $\overset{*}{\nabla}_u X \in T_a^* M$ is called the <u>contravariant differentiation at the point</u> a if*

*1). $\overset{*}{\nabla}_u X$ is linear over $\mathrm{R}$ on $u$.*

*2). $\overset{*}{\nabla}_u X$ is linear over $\mathrm{R}$ on $X$.*

*3). For any smooth function f defined in a neighbourhood of the point a*

$$\overset{*}{\nabla}_u (fX) = (uf) <X(a), \cdot > + f(a) \overset{*}{\nabla}_u X$$

**Definition.** *The contravariant differentiation is called the <u>(linear) coconnection</u>, if it is defined on an open set $U \subset M$ and for any smooth vector fields $X, Y$ on $U$ the covector field $\overset{*}{\nabla}_X Y$ is also smooth.*

**Properties** of the linear coconnection is similar to that of the linear connection.

1). *Locality*: if two vector fields are identical in some neighbourhood of the given point, then their contravariant derivatives in the direction of any vector from the tangent space in this point are equal.

2). Let $U \subset M$ be a domain, $\left\{\dfrac{\partial}{\partial x^i} \Big| i = 1,...,m\right\}$ be the basis vector fields in $U$, $\{dx^i | i = 1,...,m\}$ be the basis covector fields in $U$ generated by the same chart. Define $\partial_i := \dfrac{\partial}{\partial x^i}$, $i=1,...,m$. Introduce the Christoffel symbols for the coconnection: $\overset{*}{\nabla}_{\partial_i} \partial_j =: \sum\limits_{k=1}^{m} \Gamma_{k|ij} dx^k$ (we use tensor notations but never drop the $\Sigma$ sign). Let $X = \sum\limits_{i=1}^{m} X^i \partial_i$, $Y = \sum\limits_{j=1}^{m} Y^j \partial_j$. Then $\overset{*}{\nabla}_X Y = \sum\limits_{i,j=1}^{m} X^i (\partial_i Y^j) <\partial_j, \cdot> + \sum\limits_{i,j,k=1}^{m} X^i Y^j \Gamma_{k|ij} dx^k$.

3). Coconnection on a manifold with pseudometrics is completely defined by its Christoffel symbols. On any smooth manifold there are smooth coconnections.

4). **Definition.** *Coconnection* $\overset{*}{\nabla}$ *is called* <u>symmetric</u> *if for any smooth vector fields X,Y*   $\overset{*}{\nabla}_X Y - \overset{*}{\nabla}_Y X = <[X,Y], \cdot>$.

Coconnection is symmetric iff its Christoffel symbols are symmetric. On any smooth manifold there are smooth symmetric coconnections.

5). **Definition.** *Coconnection* $\overset{*}{\nabla}$ *is called* <u>compatible with the pseudometric, or Riemannian</u>, *if for any smooth vector fields X,Y,Z*

$$X<Y,Z> = \left(\overset{*}{\nabla}_X Y\right) \cdot Z + \left(\overset{*}{\nabla}_X Z\right) \cdot Y \quad \text{(action of a covector on a vector is written as a dot).}$$

On any manifold with a metric tensor there is the only symmetric Riemannian coconnection. It is given by the formula

$$2\left(\overset{*}{\nabla}_X Y\right) \cdot Z = (X<Y,Z> - <Y,[X,Z]>) + (Y<Z,X> - <X,[Y,Z]>) - (Z<X,Y> - <Z,[X,Y]>)$$

In the local coordinates the Christoffel symbols of this coconnection are $2\Gamma_{k|ij} = \frac{\partial g_{jk}}{\partial x^i} + \frac{\partial g_{ki}}{\partial x^j} - \frac{\partial g_{ij}}{\partial x^k}$, where $g_{ij} := <\partial_i, \partial_j>$, $i,j=1,...,m$. Since the matrix of the inner product may be degenerated, coconnection can be considered even for infinite-dimensional spaces.

In general there is no symmetric Riemannian connection on a pseudometric manifold.

**Example 7.** Consider the three-dimensional linear space with the coordinates $(x^1, x^2, x^3)$ and the metric tensor $g := Diag(1, -1-(x^3)^2, 0)$. Only the following components of the symmetric Riemannian coconnection are non-zero: $\Gamma_{3|22} = x^3$, $\Gamma_{2|32} = \Gamma_{2|23} = -x^3$. Since the metric tensor is degenerated, the Riemannian mapping $S: T_x M \to T^*_x M$ $(x \in M)$, $S(u) := <u, \cdot>$, is not an isomorphism. The image of this mapping is the set of covectors with zero third coordinate. Therefore there is no vector which is the pre-image of the covector $(\Gamma_{1|22}, \Gamma_{2|22}, \Gamma_{3|22})$ at the mapping $S$.

6). We shall need also the usual terms 'vector field along a mapping' and a **contra**variant derivative of a vector field $X:[a,b] \to TM$ *along the path* $\gamma:[a,b] \to M$:

$$\frac{\overset{*}{D}X}{dt} := \sum_{j=1}^{m} \frac{dX^j}{dt} <\partial_j, \cdot> + \sum_{i,j,k=1}^{m} \frac{d\gamma^i}{dt} X^j \Gamma_{k|ij} dx^k$$

The right part does not depend on the choice of local coordinates. This derivative has its usual properties. In particular if the coconnection is Riemannian then

$$\frac{d}{dt}<X,Y> = \frac{\overset{*}{D}X}{dt}\cdot Y + \frac{\overset{*}{D}Y}{dt}\cdot X$$ for any smooth vector fields $X,Y$ along the path $\gamma$.

### §5. Allowed coordinates and group of coordinates transformations.

**5.1. Allowed coordinates.** For manifolds with the relation of local ordering it would be advisable to introduce only such maps $h:U\to L$ that preserve local ordering. Yet there may be no such maps, i.e. this condition is too rigid [12]. Construction of such maps is in a sense equivalent to the solution of the problem 3.2. For this reason we impose more weak condition.

We saw in 2.4.5 and 2.3.2 that in linear kinematics with the indiscrete fiber (in the weakest topology) there is a naturally defined subspace $l_0$. It is the maximal linear subspace contained in the closure of the future cone of the linear kinematics. In the tangent bundle $TM$ of the manifold of the kinematic type $L$ there is a field of linear spaces $l_x:=P_x^{-1}(l_0)$, $l_x\subset T_xM$ $\forall x\in M$, i.e. a distribution. *We call a smooth atlas $\alpha$ consisting of the maps $h:U\to L$ on the manifold $M$ allowed if all maps $h\in\alpha$ preserve this distribution: $\forall x\in dom\ h\ \ d_xh(l_x)=l_0$.*

If a manifold of the kinematic type has an allowed atlas then all coordinate functions of this atlas can be numbered so that the matrices of differentials of all composition mappings $h_1\circ h_2^{-1}$ will have the blockstructure 2.3.2. If the distribution constructed above is holonomic then the allowed atlas exists. Coordinates on the appropriate integral submanifold should be chosen as coordinates with greater numbers.

There is also the case when the dimension of the future cone of the kinematics $L$ is lower than the dimension $L$. It can be treated similarly.

**5.2. The group of coordinate transformations.** Any smooth atlas $\alpha$ on the manifold $M$ such that all its maps have an additional property $\sigma$ can be extended to the smooth structure $\beta$ (the complete atlas) of all smooth in $\alpha$ maps which has the same property $\sigma$. We require (as the property $\sigma$) that each chart preserve the distribution $(l_x)_{x\in M}$ defined above, i.e. $\forall h\in\alpha\ \ \forall x\in dom\ h\ \ d_xh(l_x)=l_0$.

**Definition.** *Let M be a smooth manifold with the smooth structure β. Let each chart on M act into the linear space L. The <u>coordinate transformation</u> of the manifold M is a diffeomorphism φ:L→L such that for any chart h from the smooth structure β the composition φ∘h belong to the same smooth structure.*

In the smooth structure $β$ chosen above it is equal to the requirement $dφ(l_0)=l_0$. Therefore in any allowed atlas the Jacobi matrix of the mapping $φ$ has the blockstructure 2.3.2 or 2.3.3 depending on the ordering in *L*. Thus we are limited only by such transformations of coordinates that preserve certain geometric structure on the manifold. This geometric structure (distribution $(l_x)_{x \in M}$) is connected with the type of causality in *M*. This approach has much in common with the R.I.Pimenov's *semi-Riemannian geometry* [13].

Yet these approaches are essentially different. In homogenous spaces the structural group is acting on the space itself. This is the so-called *active* point of view on the transformations of coordinates. *Invariance with respect to action of the structural group* in homogenous spaces have another sense than it is accepted in the general relativity theory.

We use the so-called *passive* point of view on the coordinate transformations. The group of coordinate transformations defined above does not act on the tangent bundle of the manifold *M*. It acts on a smooth structure of the manifold *M,* mapping a chart to a chart ($h \mapsto φ \circ h$). This approach is closer to the term *invariance with respect to coordinate transformations* as it is accepted in the general relativity. Other discussions of various smooth structures see [15, p. 11], [16].

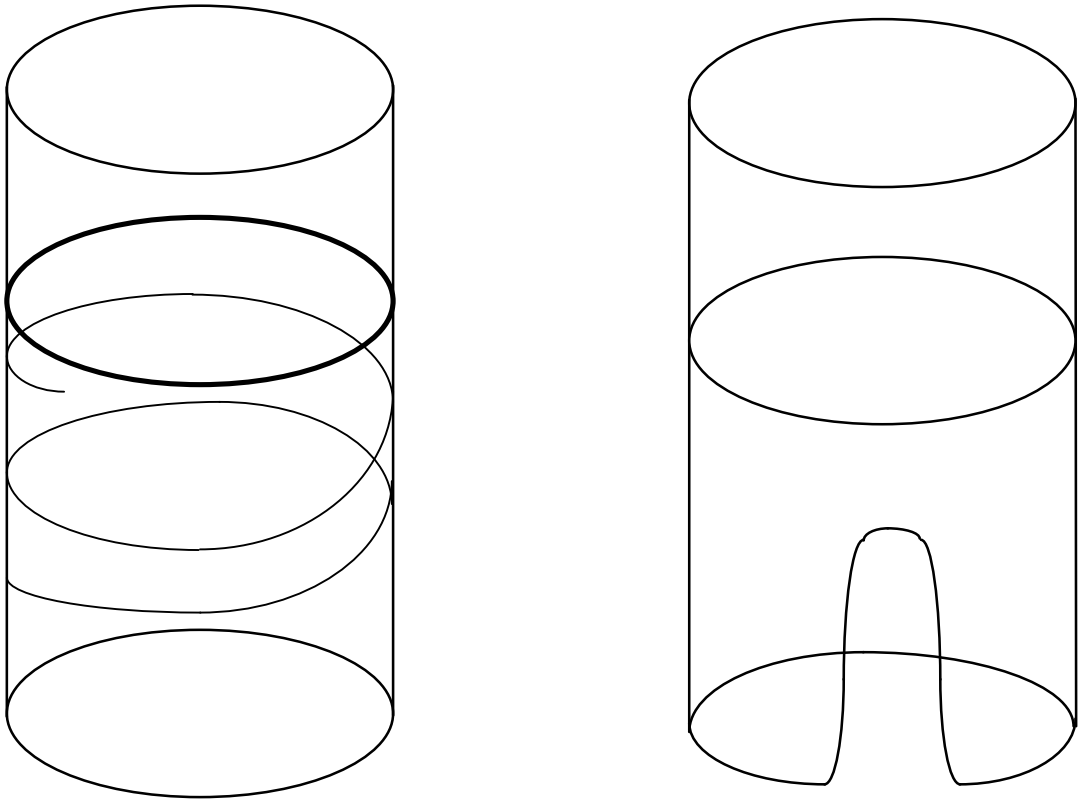

Fig. 1.